\documentclass[twoside,11pt]{article}
\pdfoutput=1
\usepackage{obs_study_style}

\usepackage{amsmath}
\usepackage{graphicx}
\usepackage{subcaption}
\usepackage[draft]{fixme}
\fxsetup{inline,nomargin,theme=color}

\def\cC{\mathcal{C}}
\def\cD{\mathcal{D}}
\def\cF{\mathcal{F}}
\def\cG{\mathcal{G}}
\def\cH{\mathcal{H}}

\def\cR{\mathcal{R}}

\def\E{\mathbb{E}}
\def\htau{\hat{\tau}}

\DeclareMathOperator*{\argmin}{arg\,min}

\newcommand\indep{\protect\mathpalette{\protect\independenT}{\perp}}
\def\independenT#1#2{\mathrel{\rlap{$#1#2$}\mkern2mu{#1#2}}}

\heading{}{}{}{}{}{Fredrik D. Johansson}

\ShortHeadings{Machine Learning Analysis of Student Mindset Interventions}{Johansson}
\firstpageno{1}

\begin{document}

\title{Machine Learning Analysis of Heterogeneity in the Effect of Student Mindset Interventions}

\author{\name Fredrik D. Johansson \email fredrikj@mit.edu \\
       \addr  Massachusetts Institute of Technology}

\maketitle
\vspace{-1em}

\begin{abstract}%
We study heterogeneity in the effect of a mindset intervention on student-level performance through an observational dataset from the National Study of Learning Mindsets (NSLM). Our analysis uses machine learning (ML) to address the following associated problems: assessing treatment group overlap and covariate balance, imputing conditional average treatment effects, and  interpreting imputed effects. By comparing several different model families we illustrate the flexibility of both off-the-shelf and purpose-built estimators. We find that the mindset intervention has a positive average effect of $0.26$, 95\%-CI $[0.22, 0.30]$, and that heterogeneity in the range of $[0.1, 0.4]$ is moderated by school-level achievement level, poverty concentration, urbanicity, and student prior expectations.
\end{abstract}

\begin{keywords}
  Machine learning, interpretability, counterfactual estimation
\end{keywords}

\vspace{1em}
\paragraph{Preface}
\emph{This work was prepared for the workshop on Empirical Investigation of Methods for Heterogeneity organized by Carlos Carvalho, Jennifer Hill, Avi Feller and Jared Murray at the 2018 Atlantic Causal Inference Conference. The workshop called for different research groups to separately analyze the same observational study. Hence, this manuscript is best read in context of the full proceedings of this workshop, to be published at a later date.}

%
% METHODOLOGY
%
\section{Methodology and Motivation}
\label{sec:method}
Machine learning (ML) has had widespread success in solving prediction problems in applications ranging from image and speech recognition~\citep{lecun2015deep} to personalized medicine~\citep{kononenko2001machine}. This makes it an attractive tool also for studying heterogeneity in causal effects. In fact, ML excels at overcoming well-known limitations of traditional methods used to solve this task. For example, matching methods struggle to perform well when confounders and covariates are high-dimensional~\citep{rubin1996matching}; generalized linear models are not flexible enough to discover variable interactions and non-linear trends; and propensity-based methods suffer from variance issues in estimation~\citep{lee2011weight}. In contrast, supervised machine learning has proven highly useful in discovering patterns in high-dimensional data~\citep{lecun2015deep}, approximating complex functions and trading off bias and variance~\citep{swaminathan2015counterfactual}.

In this observational study based on data from the National Study of Learning Mindsets (NSLM), we apply both off-the-shelf and purpose-built ML estimators to characterize heterogeneity in the effect of a student mindset intervention on future academic performance. In particular, we compare estimates of conditional average treatment effects based on linear models, random forests, gradient boosting and deep neural networks. Below, we introduce the problem of discovering treatment effect heterogeneity and describe our methodology.

%ML offers several advantages over classical methods for estimating causal effects. In particular, methods such as neural networks or regression trees learn representations that are useful for predicting the outcome, mitigating the curse of dimensionality associated with for example matching methods. They

\subsection{Problem setup}
We study the effect of a student mindset intervention based on observations of 10391 students in 76 schools from the NSLM study. The intervention, assigned at student level, is represented by a binary variable $Z \in \{0, 1\}$ and the performance outcome by a real-valued variable $Y \in \mathbb{R}$. Students are observed through covariates $S_3, C_1, C_2, C_3$ and schools through covariates $X_1, ..., X_5$\footnote{The meanings of the different covariates are described in later sections of the manuscript.}. For convenience, we let $X = [S_3, C_1, C_2, C_3, X_1, ..., X_5]^\top$ represent the full set of covariates of a student-school pair. We let $(x_{ij}, z_i, y_i)$ denote the observation  corresponding to a student $i \in \{1, ..., m\}$ in a school $j \in \{1, ..., n\}$. As each student is enrolled in at most one school, we ommit the index $j$ in the sequel. Observed treatment groups $G_0$ (control) and $G_1$ (treated) are defined by $G_z = \{i \in \{1, ..., m\} : z_i = z\}$. The full dataset of observations is denoted $\cD = \{(x_1, z_1, y_1), \ldots, (x_m, z_m, y_m)\}$, and the density of all variables $p(X,Z,Y)$.

We adopt the Neyman-Rubin causal model~\citep{rubin2005causal}, and denote by $Y(0), Y(1)$ the potential outcomes corresponding to interventions $Z=0$ and $Z=1$ respectively. The goal of this study is to characterize heterogeneity in the treatment effect $Y(1) - Y(0)$ across students and schools. As is well known, this effect is not identifiable without additional assumptions as each student is observed in only one treatment group. Instead, we estimate the \emph{conditional average treatment effect} (CATE) with respect to observed covariates $X$.
\begin{equation}
\tau(x) = \E\left[Y(1) - Y(0) \mid X = x \right]
\end{equation}
CATE is identifiable from observational data under the standard assumptions of ignorability
$$
Y(1), Y(0) \indep Z \mid X~,
$$
consistency, $Y = ZY(1) + (1-Z) Y(0)$, and overlap (positivity)
$$
\forall x : p(Z=0 \mid X=x) > 0 \Leftrightarrow p(Z=1 \mid X=x) > 0~.
$$
The CATE conditioned on the full set of covariates $X$ is the closest we get to estimating the treatment effect for an individual student. However, to design  policies, it is rarely necessary to have this level of resolution. In later sections, we estimate conditional effects also with respect to subsets or functions of $X$, such as the average effect stratified by school achievement level. By first identifying $\tau(x)$ and then marginalizing it with respect to such functions, we adjust for confounding to the best of our ability.

\subsection{Methodology overview}
The flexibility of ML estimators creates both opportunities and challenges. For example, it is typical for the number of parameters of the best performing model on a task to exceed the number of available samples. This is made possible by mitigating overfitting (high variance) through appropriate regularization. Indeed, many models achieve the best fit only after having carefully set several tuning parameters that control the bias and variance trade-off. It is standard practice in ML to use sample splitting for this purpose. Here, we apply such a pipeline to CATE estimation, proceeding through the following steps.
\begin{enumerate}
  \item Split the observed data $\cD$ into two partitions, a training set $\cD_t$ and a validation set $\cD_v$, for parameter fitting and model selection respectively.
  \item Fit estimators $f_0, f_1$ of potential outcomes $Y(0)$ and $Y(1)$ to $\cD_t$ and select tuning parameters based on held-out error on $\cD_v$
  \item Impute CATE, $\htau_i := f_1(x_i) - f_0(x_i)$, for every student in $\cD$ and fit an \emph{interpretable} model $h(x) \sim \htau$ to characterize treatment effect heterogeneity
\end{enumerate}

This pipeline allows us to find the best fitting (black-box) estimators possible in Step 2 without regard for the interpretability of their predictions. By fitting a simpler, more interpretable model to the imputed effects in Step 3, we may explain the predictions of the more complex model in terms of known quantities. This procedure is particularly well suited when the effect is a simpler function than the response and it also allows us to control the granularity at which we study heterogeneity.

The data from NSLM has a multi-level nature; students (level 1) are grouped into schools (level 2) and each level is associated with its own set of covariates. The literature is rich with studies of causal effects in multi-level settings, see for example~\citet{gelman2006data}. However, this is primarily targeted towards studying the effects of high-level (e.g. school-level) interventions on lower-level subjects (e.g. students), and the increased uncertainty that comes with such an analysis. While interventions are assigned at student-level, it is important to note that only 76 values of school-level variables are observed, which introduces the risk of overfitting to these covariates specifically. We adjust for the multi-level nature of the data in sample splitting, bootstrapping and the analysis of imputed effects.

In the following sections we describe each step of our methodology in detail.

\subsection*{Step 1. Sample splitting}%
To enable unbiased estimation of prediction error and select tuning parameters, we divide the dataset $\cD$ into two parts with 80\% of the data used for the training set $\cD_t$ and 20\% for a validation set $\cD_v$. We partition the set of schools, rather than students, making sure that the entire student body of any one school appears only in either $\cD_t$ or $\cD_v$. This is to mitigate overfitting to school-level covariates. As there are only 76 schools, random sampling may create sets that have very different characteristics. To mitigate this, we balance $\cD_t$ and $\cD_v$ by constructing a large number of splits uniformly at random and selecting the one that minimizes the Euclidean distance between summary statistics of the two sets. In particular, we compare the first and second order moments of all covariates. We increase the influence of the treatment variable $Z$ by a factor 10 in this comparison to ensure that treatment groups are split evenly across $\cD_t$ and $\cD_v$.

\subsection*{Step 2. Estimation of potential outcomes}%
The conditional average treatment effect is the difference between expected potential outcomes, here denoted $\mu_0$ and $\mu_1$. Under ignorability w.r.t. $X$ (see above), we have that
$$
\mu_z(x) := \E[Y(z) \mid X=x] = \E[Y \mid X=x, Z=z ]\;\;\mbox{ for }\;\: z\in \{0,1\},
$$
and thus, $\tau(x) = \mu_1(x) - \mu_0(x)$. A straight-forward route to estimating $\tau$ is to independently fit the conditionals $\E[Y \mid X=x, Z=z]$ for each value of $z \in \{0, 1\}$ and compute their difference. This has recently been dubbed the \emph{T-learner} approach to distinguish it from other learning paradigms~\citep{kunzel2017meta}. Below, we briefly cover theory that motivates this method and point out some of its shortcomings. To study heterogeneity, we consider several T-learners as well as two alternative approaches described below.

We approximate $\mu_0, \mu_1$ using hypotheses $f_0, f_1$ and measure their quality by the mean squared error. The group-conditional expected and empirical risks are defined as follows%
\begin{equation}
\underbrace{\vphantom{\sum_{i=1}^n}\cR_z(f_z) := \E[(\mu_z(x) - f_z(x))^2\mid Z=z]}_{\mbox{\small Expected group-conditional risk}}\;\; \mbox{ and } \;\; \underbrace{\hat{\cR}_z(f_z) := \frac{1}{|G_z|}\sum_{i \in G_z} \left( f(x_i; \theta) - y_i \right)^2}_{\mbox{\small Empirical group-conditional risk}}~.
\label{eq:risk}
\end{equation}

We never observe $\mu_z$ directly, but learn from noisy observations $y$. Statistical learning theory helps resolve this issue by bounding the expected risk in terms of its empirical counterpart and a measure of function complexity~\citep{vapnik1999overview}. For hypotheses in a class $\cF$ with a particular complexity measure $\cC_\cF(\delta,n)$ with logarithmic dependence on $n$ (e.g. a function of the covering number), it holds with probability greater than $1-\delta$ that
$$
\forall f_z \in \cF:\;\; \cR_z(f_z) \leq \hat{\cR}_z(f_z) + \frac{\cC_\cF(\delta,n)}{\sqrt{n}} - \sigma_Y^2~,
$$
where $\sigma_Y^2$ is a bound on the expected variance in $Y$ (see  \citet{johansson2018learning} for a full derivation).
This class of bounds illustrate the bias-variance trade-off that is typical for machine learning and motivates the use of regularization to control model complexity. In our experiments, we consider several T-learner models that estimate each potential outcome independently using regularized empirical risk minimization, solving the following problem.
\begin{equation}\label{eq:erm}
f_z = \argmin_{f(\cdot; \theta) \in \cF} \;\;\hat{R}_z(f(x; \theta)) + \lambda r(\theta)
\end{equation}
Here, $f(x; \theta)$ is a function parameterized by $\theta$ and $r(\theta)$ a regularizer of model parameters such as the $\ell_1$-norm (LASSO) or $\ell_2$-norm (Ridge) penalties. In our analysis, we compare four commonly used off-the-shelf machine learning estimators: ridge regression, random forests, gradient boosting and deep neural networks.

\begin{figure}[tbp!]
  \centering
  \begin{subfigure}{0.39\textwidth}
    \centering
    \includegraphics[width=\textwidth]{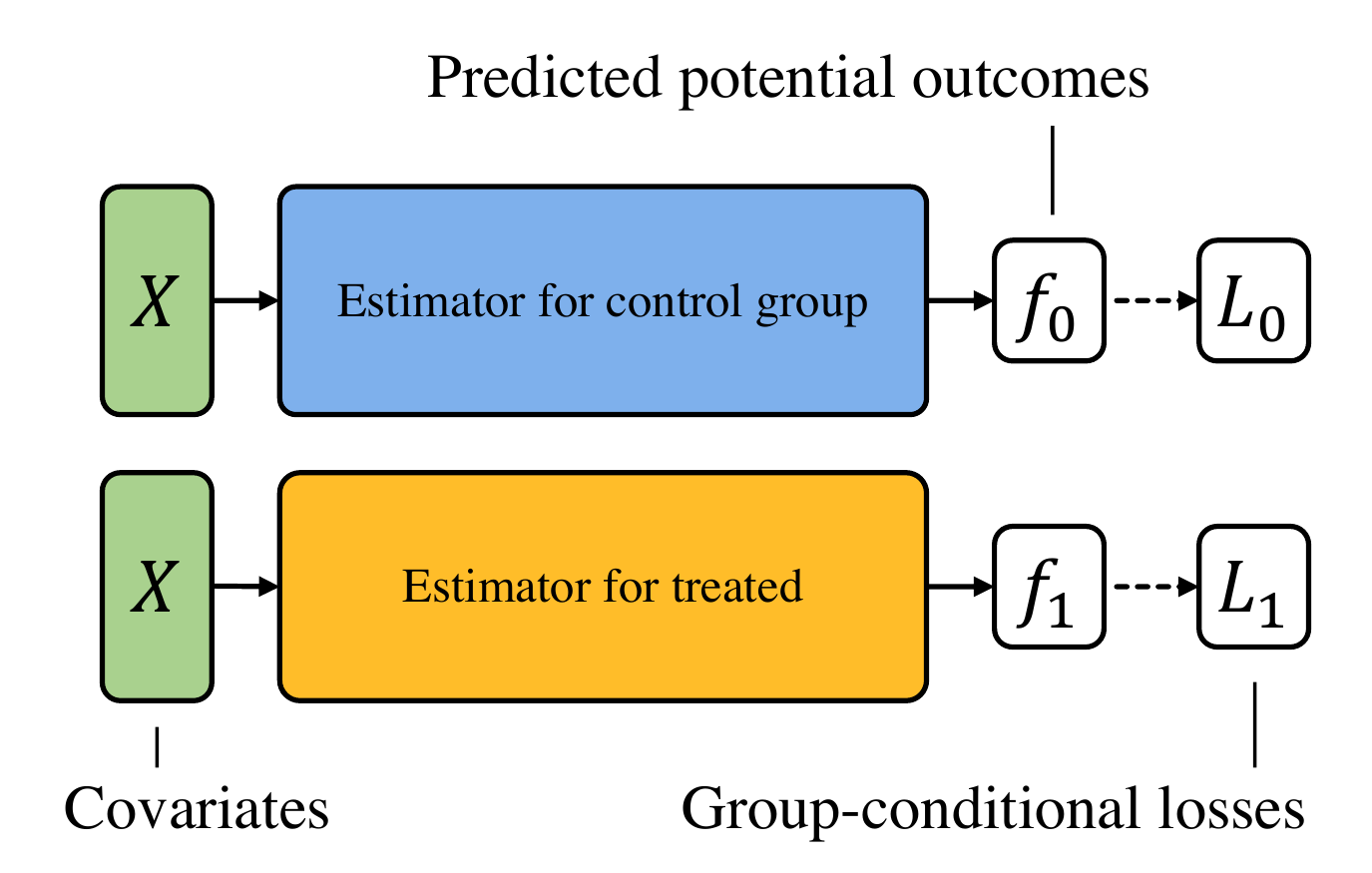}
    \caption{\label{fig:tlearner}T-learner estimator}
  \end{subfigure}
  \;
  \begin{subfigure}{0.57\textwidth}
    \centering
    \includegraphics[width=\textwidth]{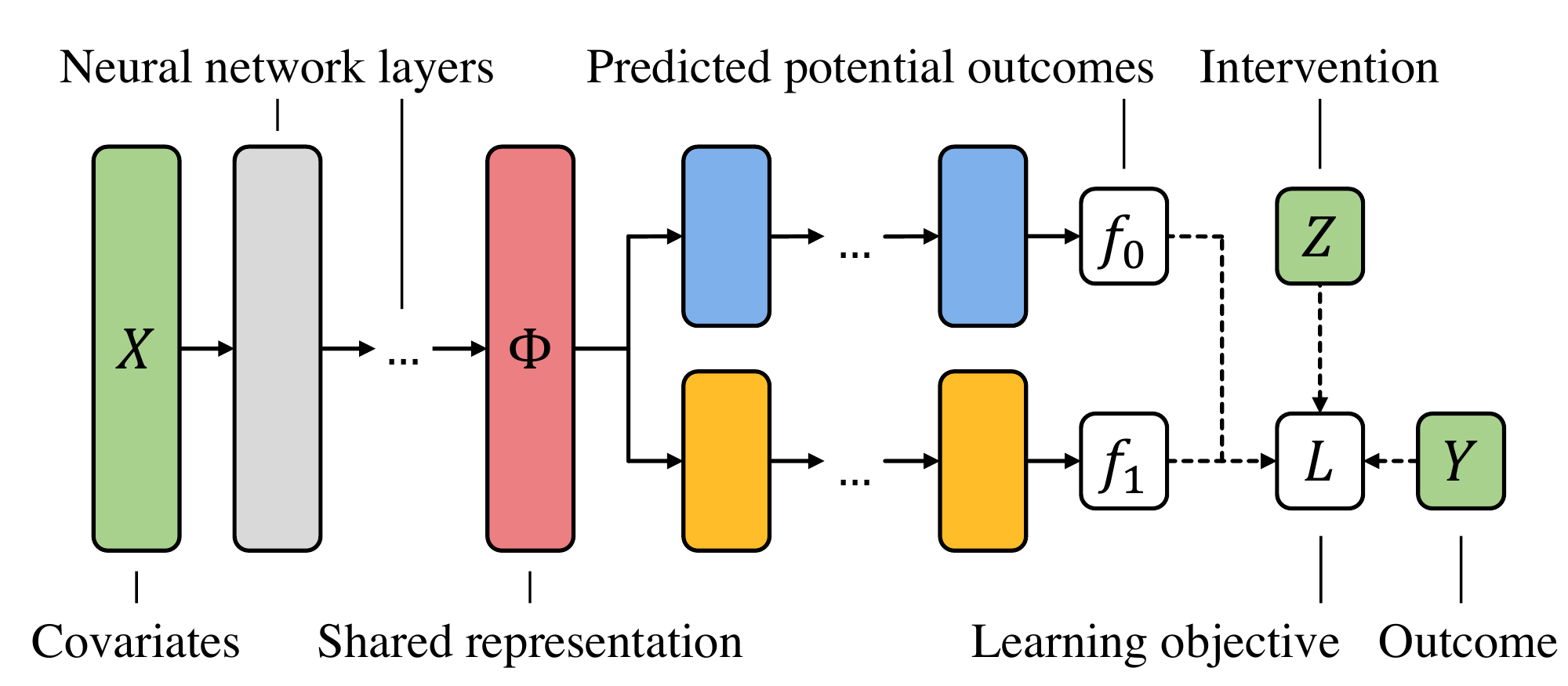}
    \caption{\label{fig:tarnet}TARNet architecture~\citep{shalit2017estimating}.}
  \end{subfigure}
  \caption{\label{fig:tlearnertarnet}Illustration of T-learner estimator (left) and TARNet architecture.}
\end{figure}

\paragraph{Sharing power between treatment groups}
A drawback of T-learners is that no information is shared between estimators of different potential outcomes. In problems where the baseline response $Y(0)$ is a more complex function than the effect $\tau$ itself, the T-learner is wasteful in terms of statistical power~\citep{kunzel2017meta,nie2017learning}. As an alternative, we apply the TARNet neural network architecture of~\citet{shalit2017estimating} which learns a representation of both treatment groups jointly, but predicts potential outcomes separately. This has the advantage of sharing information across treatment groups in learning the average response, but allows for flexibility in effect approximation. For an illustration comparing  T-learners and TARNet, see Figure~\ref{fig:tlearnertarnet}.

\subsubsection*{Generalizing across treatment groups} The careful reader may have noticed that the population and empirical risk in equations  \eqref{eq:risk}--\eqref{eq:erm} are defined with respect to the observed treatment assignments. To estimate CATE, we want our estimates of potential outcomes to be accurate for the counterfactual assignment as well. In other words, we want for the risk on the full cohort,
$$R(f_z) := \E[(\mu_z(x) - f_z(x))^2]$$
to be small. When treatment groups $p(X \mid Z=0)$ and $p(X \mid Z=1)$  are highly imbalanced, the expected risk within one treatment group may not be representative of the risk on the full population. This is another drawback of T-learner estimators, which do not adjust for this discrepancy.

In recent work, \citet{shalit2017estimating} characterize the difference between $R(f_z)$ and $R_z(f_z)$ and bound the error in CATE estimates using a  distance metric between treatment groups. In particular, they considered the \emph{integral probability metric} (IPM) family of distances, resulting in the following relation between full-cohort and treatment group risk.
\begin{equation}\label{eq:ipmbound}
R(f_z) \leq R_z(f_z) + \mbox{IPM}_{\cG}(p(X\mid Z=0), p(X\mid Z=1))
\end{equation}
In~\citet{shalit2017estimating}, the authors used the kernel-based Maximum Mean Discrepancy~\citep{gretton2012kernel} to regularize and \emph{balance} the representations learned by the TARNet architecture, minimizing an upper bound on the CATE risk. We apply this approach, called Counterfactual Regression (CFR), in our analysis.

\subsection*{Step 3. Characterization of heterogeneity in CATE}%
After fitting models $f_0, f_1$ for each potential outcome, the conditional average treatment effect is imputed for each student by $\htau_i = f_1(x_i) - f_0(x_i)$. Unlike with linear regressors, the predictions of most ML estimators are difficult to interpret directly through model parameters. For this reason, ML models are often considered \emph{black-box} methods~\citep{lipton2016mythos}. However, in the study of heterogeneity, it is crucial to characterize for which subjects the effect of an intervention is low and for which it is high. To accomplish this, we adopt the common practice of post-hoc interpretation---fitting a simpler, more interpretable model $h \in \cH$ to the imputed effects $\{\htau_i\}$.

In its very simplest form $h(x_i)$ may be a function of a single attribute, such as the school size, effectively averaging over other attributes. This is usually a good way of discovering global trends in the data but will neglect meaningful interactions between variables, much like a linear model. As a more flexible alternative, we also fit decision tree models and inspect the learned trees. Trees of only two variables may be visualized directly in the covariate space, and larger trees in terms of their decision rules.

%
% WORKSHOP RESULTS
%
\section{Workshop results}
\label{sec:workshop}
We present the first results of our analysis as shown in the workshop Empirical Investigation of Methods for Heterogeneity at the Atlantic Causal Inference Conference, 2018.

\subsection{Covariate balance}
\begin{figure}[tbp!]%
  \centering
  \begin{subfigure}{.52\textwidth}
    \includegraphics[width=\textwidth]{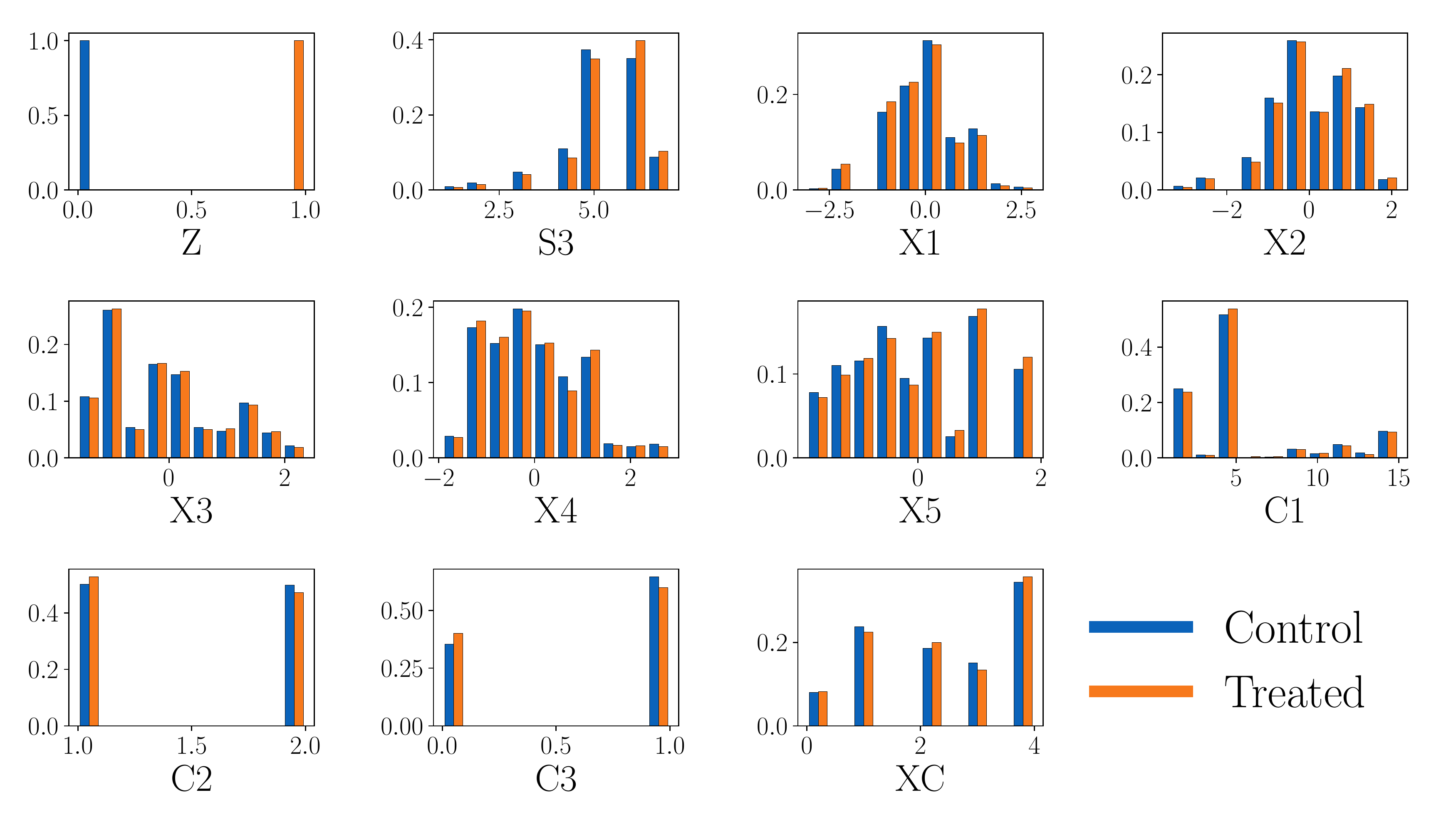}
    \caption{\label{fig:overlap_marg} Marginals of covariates $X$}
  \end{subfigure}
  %\begin{subfigure}{.48\textwidth}
  %  \includegraphics[width=\textwidth]{fig/overlap_pca.pdf}
  %  \caption{\label{fig:overlap_pca} PCA projection of covariates $X$}
  %\end{subfigure}
  \;
  \begin{subfigure}{.44\textwidth}
    \includegraphics[width=\textwidth]{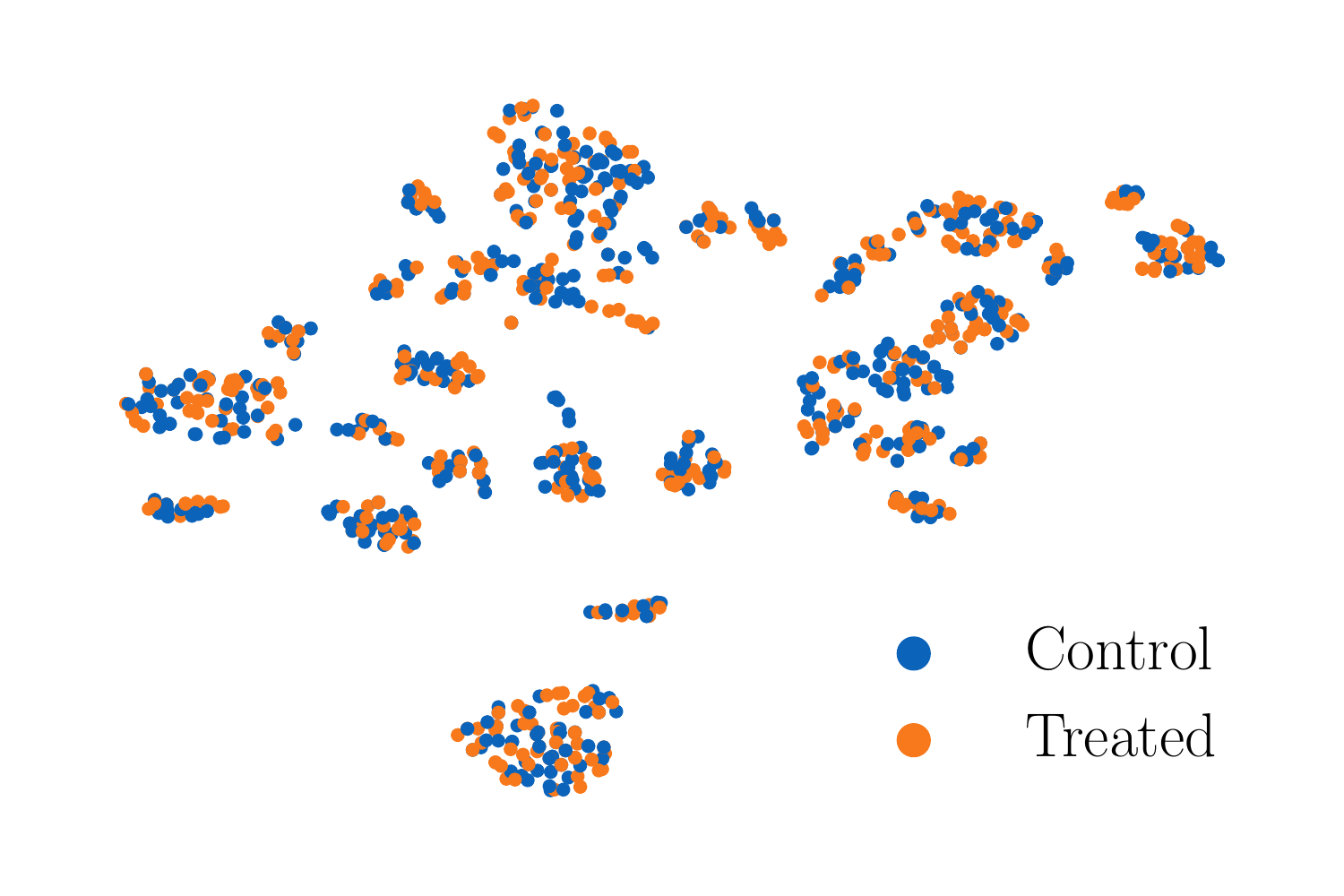}
    \caption{\label{fig:overlap_tsne} t-SNE projection of covariates $X$}
  \end{subfigure}
  \caption{\label{fig:overlap}Examination of overlap through covariate marginal distributions and a low-dimensional t-SNE projection of covariates~\citep{maaten2008visualizing}. Marker color corresponds to treatment assignment $Z$. Best viewed in color. }
\end{figure}

First, we investigate the extent to which the overlap assumption holds true by comparing the covariate statistics of the treatment and control groups. In Figure~\ref{fig:overlap}, we visualize the marginal distributions of each covariate, as well as a 2D t-SNE projection of the entire covariate set~\citep{maaten2008visualizing}.
The observed difference between the marginal covariate distributions of the two treatment groups is very small. Also the non-linear t-SNE projection reveals little difference between treatment groups. The less imbalance between treatment groups, the closer our problem is to standard supervised learning. Said differently, the density ratio $p(Z=1 \mid X)/p(Z=0 \mid X)$ is close to $1.0$ and the IPM distance between conditional distributions, see \eqref{eq:ipmbound}, is small. Hence, we expect neither propensity re-weighting nor balanced representations (e.g. CFR) to have a large effect on the results.

\subsection{Estimation of potential outcomes}
\begin{figure}[tbp!]
  \centering
  \begin{subtable}{.55\textwidth}
    \centering
    \renewcommand*{\arraystretch}{1.1}
    \small
    \begin{tabular}{lll}
      %\bf Estimator & \bf MSE & \bf $\mathbf{R^2}$ \\
      \bf Estimator & \multicolumn{1}{c}{$\hat{\mbox{\bf ATE}}$} & \multicolumn{1}{c}{$\mathbf{R^2}$} \\
      \hline
      %Ridge regression & $0.29 \pm x.x $ & $0.28 \pm x.x$  \\
      %Random forest & $0.29 \pm x.x$ & $0.28 \pm x.x$  \\
      %Neural network & $0.33 \pm x.x$ & $0.18 \pm x.x$  \\
      %TARNet & $0.29 \pm x.x$ & $0.30 \pm x.x$  \\
      %CFR & $0.29 \pm x.x$ & $0.30 \pm x.x$  \\
      Na\"ive & $0.30 \;\;\;\;\;\;\;\;\; $ & \multicolumn{1}{c}{---\;} \\
      \hline
      RR & $0.26\;\; [0.22, 0.29]$ & $0.26\;\; [0.17, 0.29]$  \\
      RF & $0.27\;\; [0.23, 0.30]$ & $0.25\;\; [0.22, 0.29]$  \\
      GB & $0.26\;\; [0.21, 0.30]$ & $0.25\;\; [0.20, 0.30]$  \\
      NN & $0.27\;\; [0.17, 0.38]$ & $0.14\;\; [-0.08, 0.23]$  \\
      \hline
      TARNet & $0.26\;\; [0.23, 0.30]$ & $0.27\;\; [0.21, 0.31]$  \\
      CFR & $0.26\;\; [0.22, 0.30]$ & $0.27\;\; [0.22, 0.31]$  \\
    \end{tabular}
    \vspace{.5em}
    \caption{ATE and held-out $R^2$ score with 95\% school-level cluster bootstrap confidence intervals.}
  \end{subtable}
  ~
  \begin{subfigure}{.42\textwidth}
    \centering
    \includegraphics[width=.95\textwidth]{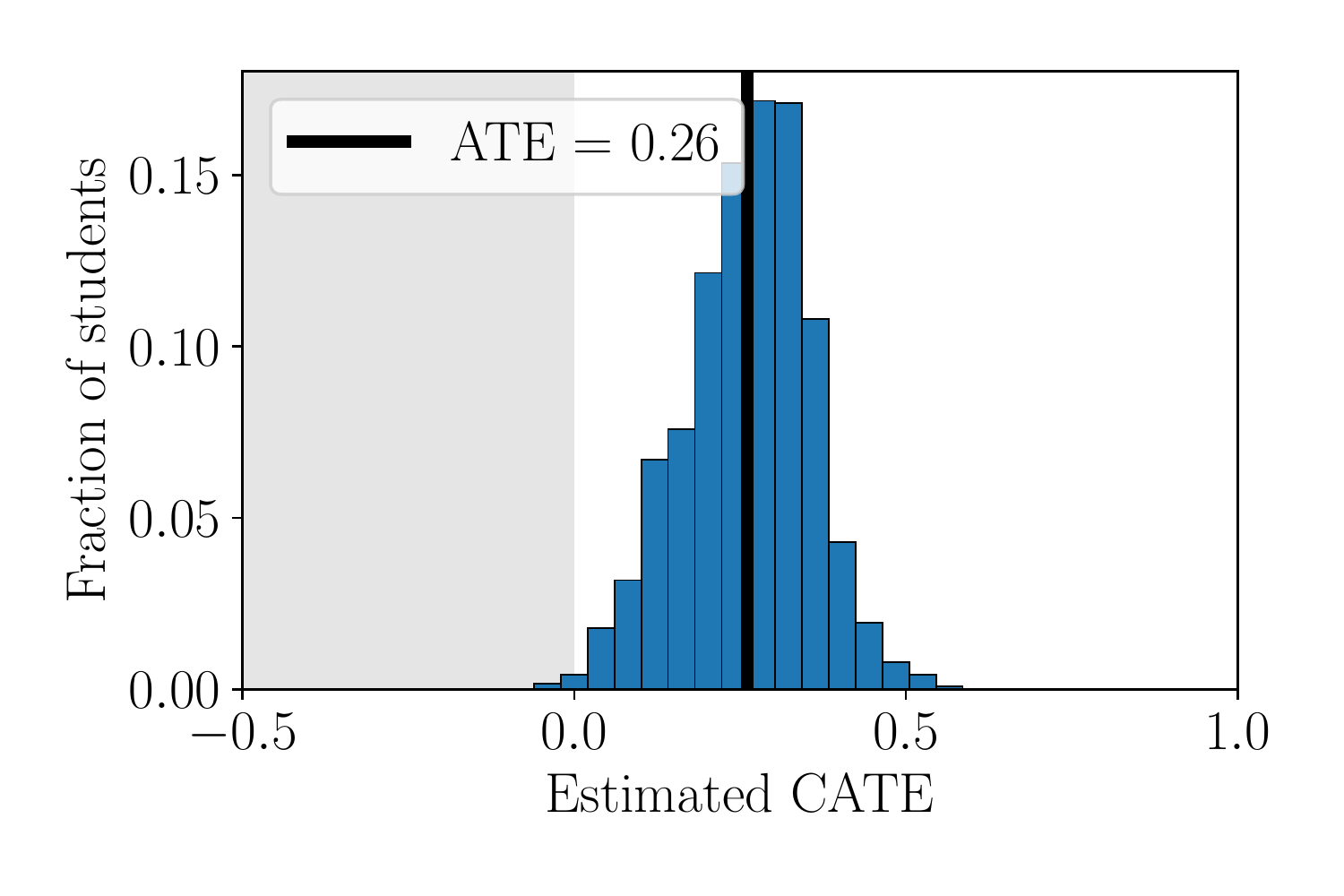}
    \caption{\label{fig:cate_hist} Histogram of CATE (CFR).}
  \end{subfigure}
  %\begin{subfigure}{.45\textwidth}
  %  \includegraphics[width=\textwidth]{fig/potential_outcomes.pdf}
  %  \caption{\label{fig:pot_out} \fxfatal{Caption!}}
  %\end{subfigure}
  \caption{\label{tbl:pot}Comparison between the na\"ive estimator, T-learners, and representation learning methods (left). Heterogeneity in treatment effect estimated by CFR (right).}
\end{figure}
In our analysis, we compare four T-learners based on ridge regression (RR), random forests (RF), gradient boosting (GB), and neural networks (NN). In addition, we compare the representation-learning algorithms TARNet and Counterfactual Regression (CFR)~\citep{shalit2017estimating}. For each estimator family, we fit models of both potential outcomes on the training set $\cD_t$ and select tuning parameters based on the held-out $R^2$ score on the validation set $\cD_v$. To estimate  uncertainty in model predictions, we perform school-level bootstrapping of the training set~\citep{cameron2008bootstrap}, fitting each model to each bootstrap sample\footnote{The bootstrap analysis was added after the workshop results, but is presented here for completeness.}. In Table~\ref{tbl:pot}, we give the estimate of the average treatment effect (ATE) from each model, the held-out $R^2$ score of the fit of factual outcomes, and 95\% confidence intervals based on the empirical bootstrap. In addition, we give the na\"{i}ve estimate of the ATE---the difference between observed average outcomes in the two treatment groups.

We see that all methods produce very similar estimates of ATE and perform comparably in terms of $R^2$. As expected, based on the small covariate imbalance shown in the previous section, the regression adjusted estimates are close to the na\"ive estimate of the ATE. This likely also explains the small difference between TARNet and CFR, as even for moderate to large imbalance regularization, the empirical risk dominates the objective function. The performance of the neural network T-learner would likely be improved with a different choice of architecture or tuning parameters. This is consistent with  \citet{shalit2017estimating} in which TARNet architecture achieved half of the error of the T-learner on the IHDP benchmark.

%
% Heterogeneity
%
\subsection{Heterogeneity in causal effect}
We examine further the CATE for each student imputed by the best fitting model. As CFR had a slight edge in $R^2$ over T-learning estimators (although confidence intervals overlap) and has stronger theoretical justification, we analyze the effects imputed by CFR below. In Figure~\ref{fig:cate_hist}, we visualize the distribution of imputed CATEs. We see that for almost all students, the effect is estimated to be positive, indicating an improvement in performance as an effect of the mindset intervention. Recall that the average treatment effect was estimated to be $0.26$. Around 95\% of students were estimated to have an effect in the range $[0.05, 0.45]$. For reference, the mean of the observed outcome was $0.10$ and the standard deviation $0.64$.

\begin{figure}[tbp!]%
  \centering
  \includegraphics[width=0.8\textwidth]{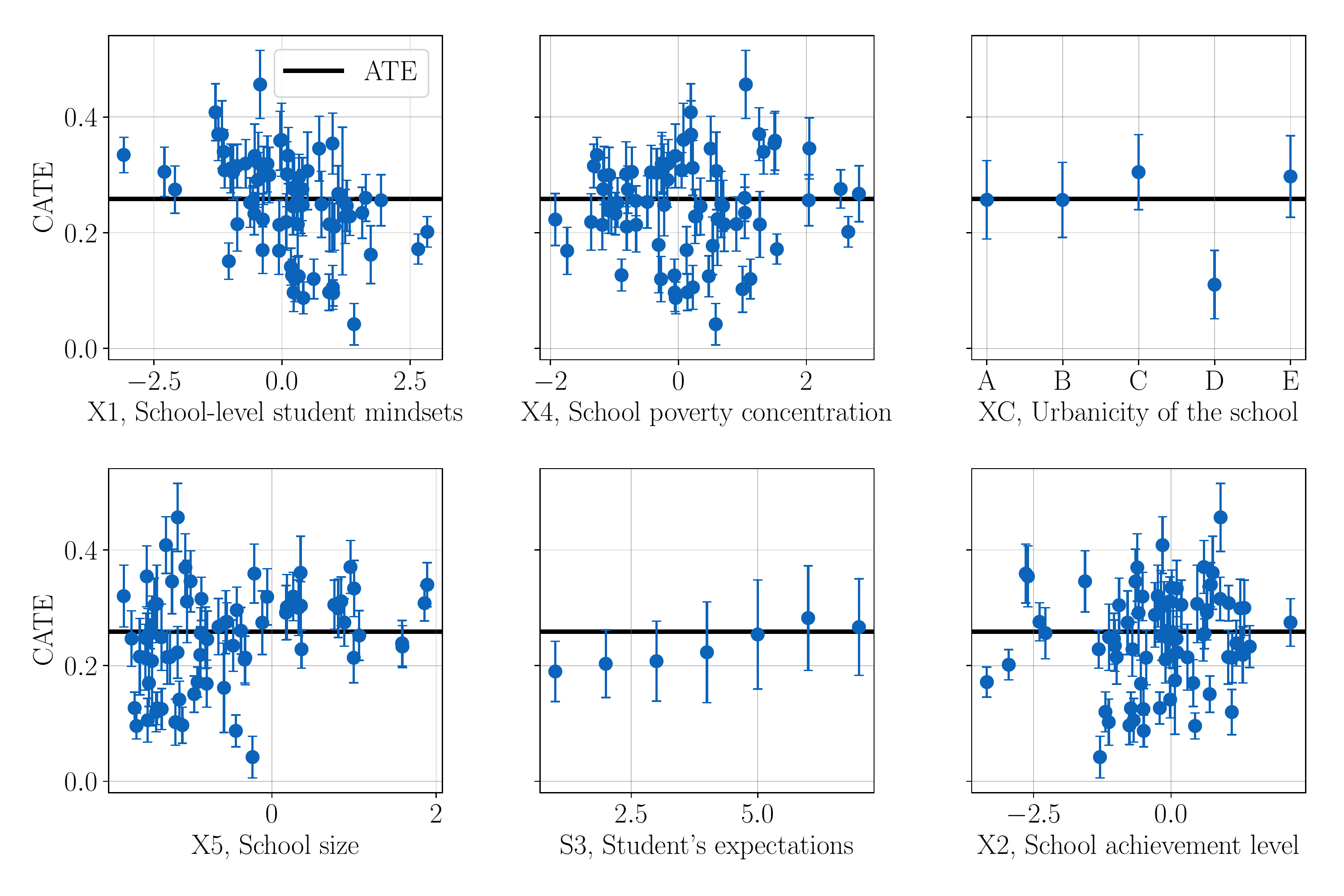}
  \caption{\label{fig:covariate_strat}Heterogeneity in causal effect estimated using counterfactual regression (CFR) stratified by different covariates. Bars indicate variation in point estimates across subjects. }
\end{figure}

To discover drivers of heterogeneity we fit a random forest estimator to imputed effects and inspect the feature importance of each variable---the frequency with which it is used to split the nodes of a tree. The five most important variables of the random forest were $X_1$, $X_4$, $X_C$, $X_5$ and $S_3$. In Figure~\ref{fig:covariate_strat} we stratify imputed CATE with respect to these variables, as well as $X_2$ which is of interest to the study organizers. We see a strong trend that the effect of the intervention decreases with prior school-level student mindset, $X_1$. The urbanicity of the school, $X_C$, is a categorical variable for which Category D appears to be associated with substantially lower effect. In contrast, the effect of the intervention appears to increase with students' prior expectations $S_3$. One of the questions of the original study was whether there exists a ``Goldilocks effect'' for school-achievement level $X_2$, meaning that the intervention only has an effect for schools that are neither achieving too poorly nor too well. These results cannot confirm this hypothesis, nor reject it.

%
% POST-WORKSHOP
%
\section{Post-workshop analysis}
\label{sec:postworkshop}
Heterogeneity in treatment effect may be a non-linear or non-additive function of observed covariates. Such patterns remain hidden when analyzing CATE as a function of a single variable at a time or using linear regression. To reveal richer patterns of heterogeneity, we fit highly regularized regression tree models and inspect their decision rules. First, we consider combinations only of pairs of variables at a time. We note that for school-level variables, only 76 unique values exist, one for each school. To prevent overfitting to these variables, we require that each leaf in the regression tree contains samples from at least 10 schools. When student-level covariates are included, we require leaves to have samples of at least 1000 students.

\begin{figure}[tbp!]%
  \centering
  \begin{subfigure}{.46\textwidth}
    \includegraphics[width=\textwidth]{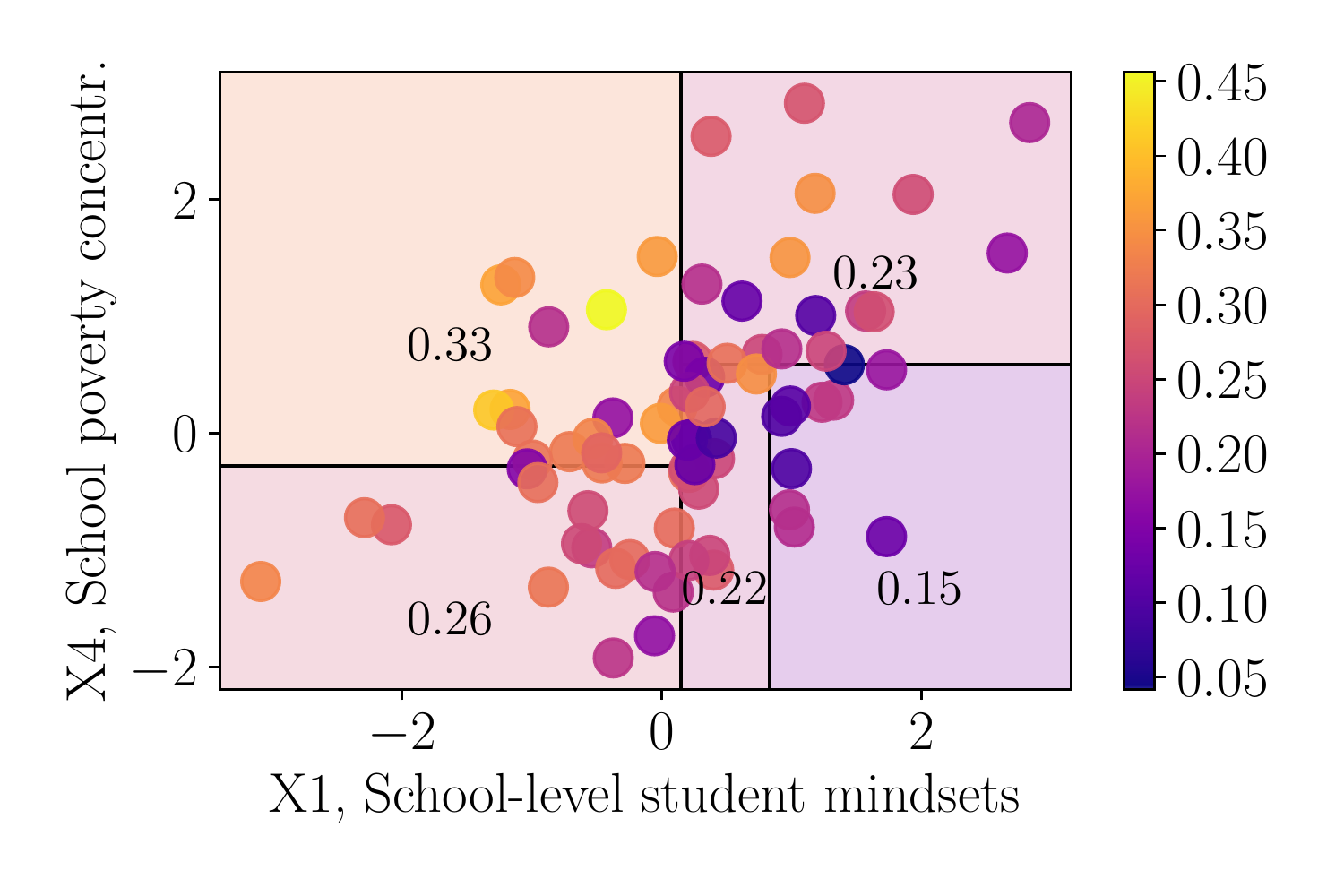}
    \caption{\label{fig:cate_x1_x2} CATE vs. X1 and X4 (school-level)}
  \end{subfigure}
  \;
  \begin{subfigure}{.46\textwidth}
    \includegraphics[width=\textwidth]{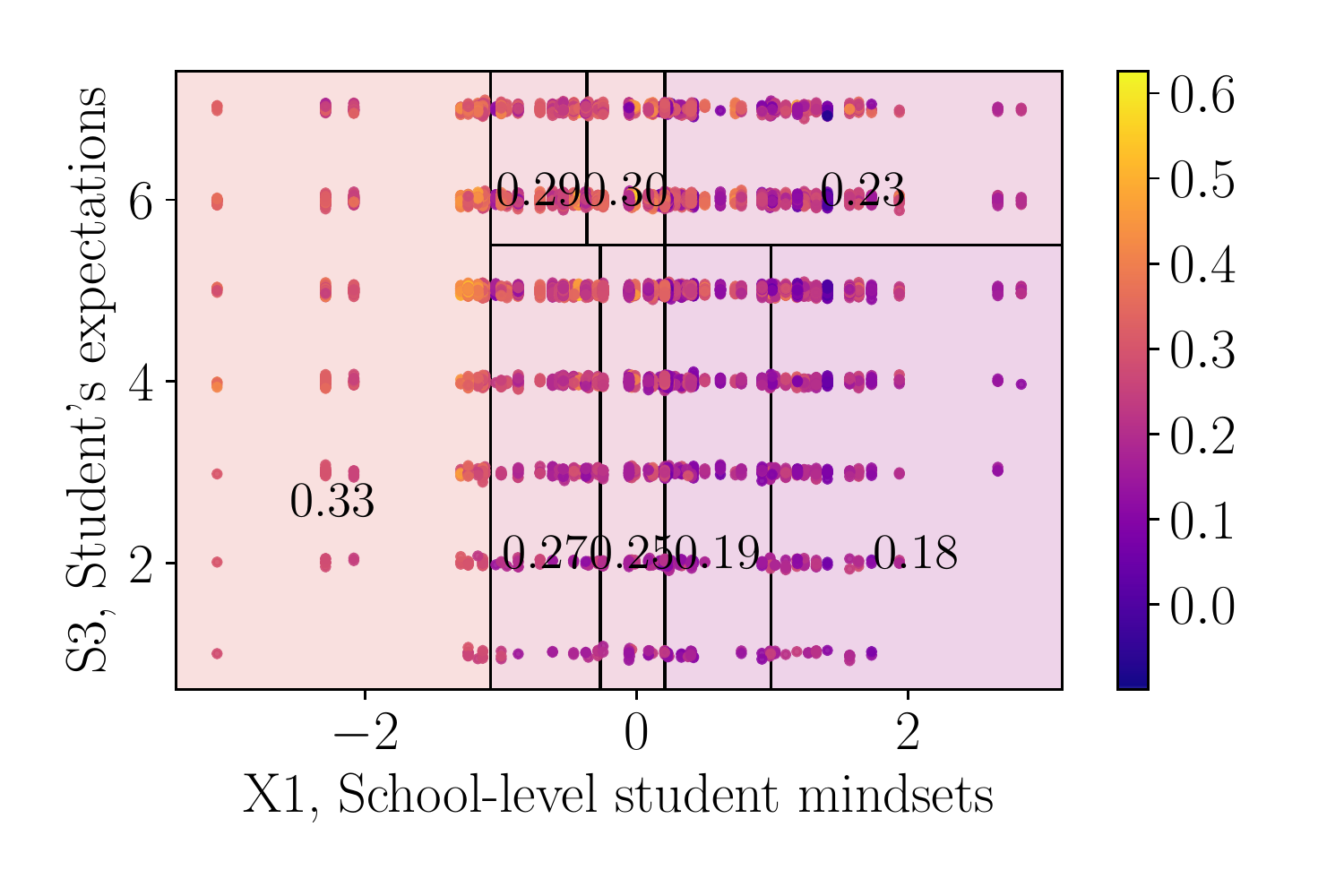}
    \caption{\label{fig:cate_x1_s2} CATE vs. X1 and S3 (student-level)}
  \end{subfigure}
  \caption{\label{fig:cate_tree}Intrepretation of CATE estimates using regression trees fit to pairs of covariates. Each dot represents a single school (left) or student (right). The color represents the predicted CATE. Black lines correspond to leaf boundaries. Background color and numbers in boxes correspond to the average predicted CATE in that box. Best viewed in color.}
\end{figure}

In Figure~\ref{fig:cate_tree}, we visualize trees fit to two distinct variable pairs. We note a very slight non-linear pattern in heterogeneity as a function of $X_1$ (school-level student mindset) and $X_4$ (school poverty concentration), and that $X_1$ explains a lot of the variance observed at moderate values of $X_4$ in Figure~\ref{fig:covariate_strat}. We emphasize, however, that the sample size at the school-level is small, and that observed patterns have high variance. In the right-hand figure, $S_3$ (student's expectations) appears associated with a larger effect only if the average mindset of the school is sufficiently high. This pattern disappears when using a linear model. In the Appendix, we show a regression tree fit to the entire covariate set.

%
% DISCUSSION
%
\section{Discussion}
\label{sec:discussion}

Machine learning offers a broad range of tools for flexible function approximation and provides theoretical guarantees for statistical estimation under model misspecification. This makes it a suitable framework for estimation of causal effects from non-linear, imbalanced or high-dimensional observational data. The flexibility of machine learning comes at a price however: many methods come with tuning parameters that are challenging to set for causal estimation; models are often difficult to optimize globally; and interpretability of models suffers. While progress has been made independently on each of these problems, a standardized set of tools has yet to emerge.

In the analysis of the NLSM data, machine learning appears well-suited to study overlap, potential outcomes and heterogeneity in imputed effects. However, the analysis also opens some methodological questions. The multi-level nature of covariates is not accounted for in most off-the-shelf ML models and regularization of models applied to multi-level data has been comparatively less studied than for single-level data. In addition, as pointed out by several authors~\citep{kunzel2017meta,nie2017learning}, the T-learner approach to causal effect estimation may suffer from compounding bias and from wasting statistical power. This may be one of the reasons we observe a slight advantage of representation learning methods such as TARNet and CFR.

\vskip 0.2in
\bibliography{sample}

\clearpage
\appendix
\section{Regression tree explanation of CATE}
\begin{figure}[h!]
  \centering
  \includegraphics[width=\textwidth]{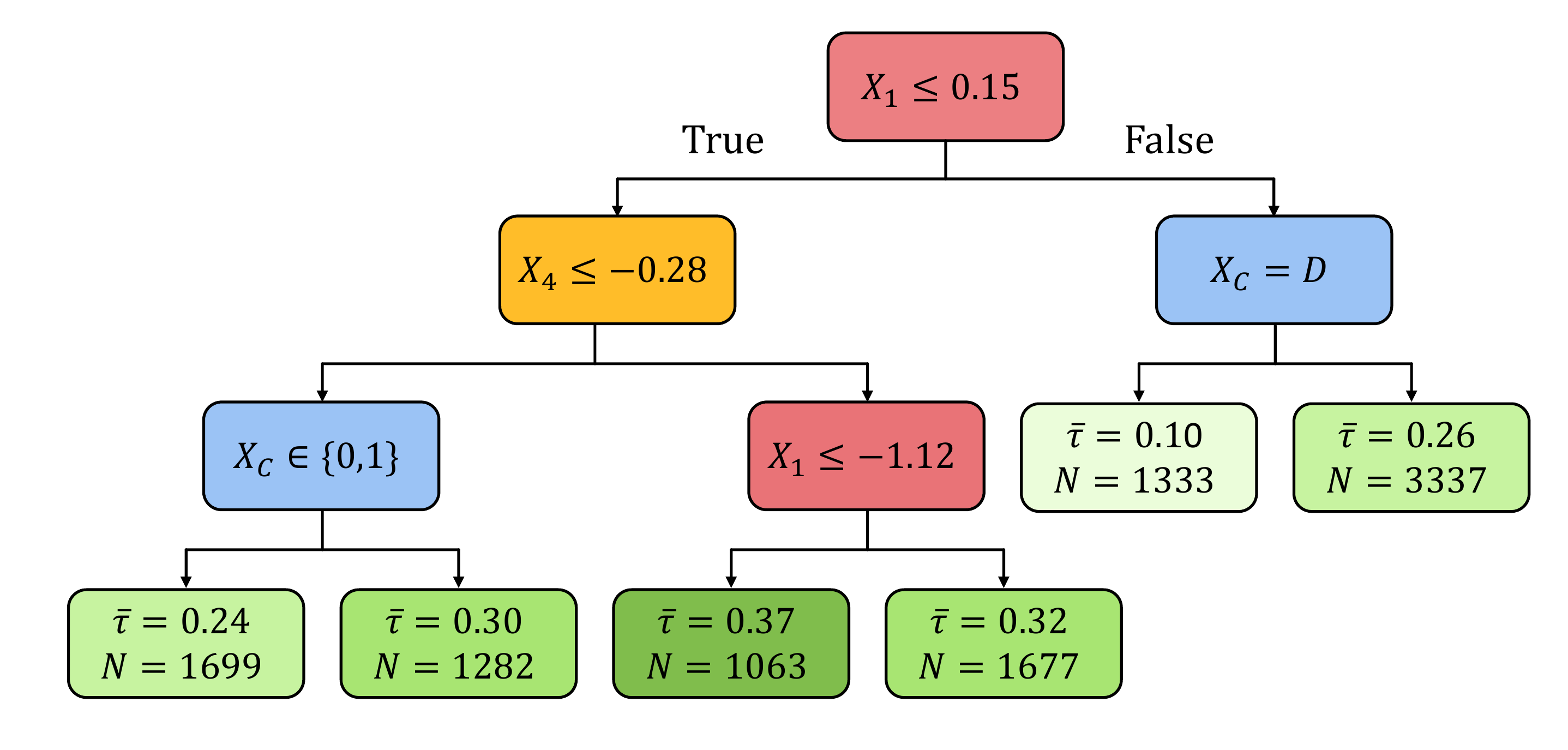}
  \caption{\label{fig:dectree}Visualization of a regression tree fit to the imputed CATE values based on the full covariate set.}
\end{figure}

\end{document}